\documentclass[twocolumn,showpacs,amsmath,amssymb,prd]{revtex4}

\usepackage{epsfig,epsf,graphics,psfrag}
\usepackage{times}
\usepackage{float}
\usepackage{color}
\usepackage{amsfonts,amssymb,stmaryrd,latexsym,amsmath}
\usepackage{multirow}
\usepackage{array} 
\usepackage{mathrsfs}
\usepackage{booktabs}
\usepackage{enumerate}
\usepackage{color}
\usepackage{subfigure}
\usepackage{hhline}
\usepackage[hyperindex=true]{hyperref}
\usepackage{epstopdf}

\usepackage{slashed}

\newcommand{\lamPKpi}{$\Lambda^+_c\to p K^- \pi^+$}

\begin{document}

\title{Visible narrow cusp structure in $\Lambda_c^+\to p K^- \pi^+$ enhanced by triangle singularity}

\author{Xiao-Hai~Liu$^{1}$}\email{xiaohai.liu@tju.edu.cn}
\author{Gang Li$^{2,4}$}\email{gli@qfnu.edu.cn}
\author{Ju-Jun Xie$^{3,5,6}$}\email{xiejujun@impcas.ac.cn}
\author{Qiang Zhao$^{4,5,7}$}\email{zhaoq@ihep.ac.cn}

\affiliation{
$^1$Center for Joint Quantum Studies and Department of Physics, School of Science, Tianjin University, Tianjin 300350, China\\
$^2$School of Physics and Engineering, Qufu Normal
University, Shandong 273165, China\\
$^3$Institute of Modern Physics, Chinese Academy of
Sciences, Lanzhou 730000, China\\
$^4$ Institute of High Energy Physics and Theoretical Physics Center for Science Facilities,
Chinese Academy of Sciences, Beijing 100049, China \\
$^5$ School of Nuclear Science and Technology, University of Chinese Academy of Sciences, Beijing 100049, China\\
$^6$School of Physics and Engineering, Zhengzhou University, Zhengzhou, Henan 450001, China \\
$^7$ Synergetic Innovation Center for Quantum Effects and
Applications (SICQEA), Hunan Normal University, Changsha 410081,
China }

\date{\today}

\begin{abstract}

A resonance-like structure as narrow as 10 MeV is observed in the
$K^-p$ invariant mass distributions in $\Lambda_c^+\to p K^- \pi^+$
at Belle. Based on the large data sample of about 1.5 million events
and the small bin width of just 1 MeV for the $K^-p$ invariant mass
spectrum, the narrow peak is found precisely lying at the
$\Lambda\eta$ threshold. While lacking evidence for a quark model
state with such a narrow width at this mass region, we find that
this narrow structure can be naturally identified as a threshold
cusp but enhanced by the nearby triangle singularity via the
$\Lambda$-$a_0(980)^+$ or $\eta$-$\Sigma(1660)^+$ rescatterings.

\end{abstract}
 \maketitle

\section{Introduction}

The discontinuation of the scattering amplitude caused by an $S$-wave open threshold generally will result in the cusp phenomenon in the energy spectrum. However, although the cusp phenomena were noticed long time ago and have been discussed broadly in the literature, it is very difficult to measure them experimentally. One reason is that the cusp can only occur at threshold, and to observe it requires an energy scan in the vicinity of the threshold. This imposes a challenge on the detector for a high performance in energy resolution. Besides, the cusp structure is generally much less prominent over the background than a pole structure. Thus, a huge data sample is necessary for isolating the signal out of complicated background.
One classical example is the $\pi\pi$ scattering where
the charge-exchange reaction $\pi^+\pi^-\to \pi^0\pi^0$ can produce a cusp in the $\pi^0\pi^0$ invariant mass spectrum at the $\pi^+\pi^-$ threshold \cite{Batley:2005ax,Budini:1961bac,Cabibbo:2004gq,Colangelo:2006va,Bissegger:2007yq}.
This observation is based on a data sample of $2.287\times 10^7$ events for the $K^\pm\to \pi^\pm\pi^0\pi^0$ decays, and the excellent energy resolution for the $\pi^0\pi^0$ invariant mass spectrum~\cite{Batley:2005ax}. This measurement provided a precise determination of the $\pi\pi$ scattering length, which was suggested by Cabibbo~\cite{Cabibbo:2004gq}. Similar proposals were also suggested for some other precise experiments~\cite{Kubis:2009sb,Liu:2012dv}.

There are also some other less prominent cusps observed in
experiments, such as the one in $\gamma p\to \pi^0 p$ at the $\pi^+
n$ threshold \cite{Bernstein:1996vf}. In recent years, the cusp
phenomena have ever been introduced to describe some resonance-like
structures in both the heavy hadron
~\cite{Chen:2011pv,Bugg:2011jr,Swanson:2014tra} and light hadron
sectors~\cite{Oller:1997ti,Kornicer:2016axs,Oller:2000fj,Jido:2003cb,Oller:2006jw,Guo:2012vv,Roca:2013cca,Wang:2015qta,liu:2017efp,Khemchandani:2018amu,Xie:2018gbi}.
But it should be warned that depending on the coupling strength to
the open threshold, not all cusp effects would produce predominant
resonance-like enhancements~\cite{Guo:2014iya}. In fact, in most
cases the cusp structures are tiny kinks that can hardly be
identified without sufficiently large high-quality data samples.

The Belle collaboration recently reported a narrow structure
observed in the $K^-p$ invariant mass spectrum in \lamPKpi, of which
the mass is 1663 MeV, and the width is 10 MeV
\cite{shen:workshop,Gabyshev:workshop}. The signal yields of
\lamPKpi decays at Belle is about $1.452\times 10^6$ and the bin
width of $K^-p$ invariant mass is only 1 MeV, which means this is a
very precise measurement. From the latest \textit{Particle Data
Group} (PDG)~\cite{Tanabashi:2018oca}, there are a few hyperon
resonances whose masses are close to 1663 MeV, such as
$\Lambda(1670)$, $\Lambda(1690)$, $\Sigma(1660)$, and
$\Sigma(1670)$, but all their widths are much larger than 10 MeV.
Considering the mass spectra of hyperon resonances and their
couplings to open channels, we see that none of those established
hyperons can account for such a narrow structure. Interestingly, one
notices that the peak position of the structure in $\Lambda^+_c \to
p K^- \pi^+$ is coincident with the $\Lambda \eta$ mass threshold
$\sim 1663.5\ \mbox{MeV}$. This could provide an important clue for
understanding the narrow structure that may strongly be correlated
with the $\Lambda\eta$ threshold cusp. Meanwhile, as mentioned
earlier, the two-body unitarity cut usually cannot lead to such a
narrow peak~\cite{Guo:2014iya}. In Ref.~\cite{Miyahara:2015cja}, a
chiral unitary approach is employed to study the
$\bar{K}N-\pi\Sigma-\eta\Lambda$ coupled channel interactions and it
was shown that the two-body unitarity cut would not produce narrow
structures at the $\Lambda\eta$ threshold in $\Lambda^+_c \to p K^-
\pi^+$. Therefore, a detailed analysis of the analytical property of
the transition amplitude taking into account the $\Lambda\eta$ open
threshold, but looking at more leading contributions, could be the
key for unlocking the puzzle about the narrow structure.

Notice that the three-body decays \lamPKpi could receive
contributions from rescattering processes, e.g. via the
Cabibbo-favored intermediate processes $\Lambda_c^+\to \Lambda
a_0(980)^+$ and $\Lambda_c^+\to \eta \Sigma^{*+}$ (here,
$\Sigma^{*+}$ representing an excited hyperon with $I=1$), we find
that the $\Lambda\eta$ rescatterings are located in the vicinity of
the so-called ``triangle singularity (TS)" kinematic region. As the
leading singularity of the complex scattering amplitude, its
association with the two-body cut near the physical boundary will
strongly enhance the two-body cusp effects. We find this mechanism
can provide a natural explanation for the narrow cusp structure
observed in \lamPKpi. It should be noted that the TS mechanism has
been recognized recently to play a crucial role in the understanding
a lot of puzzling threshold phenomena in experiment. Some relevant
topical discussions can be found in
Refs.~\cite{Wu:2011yx,Wang:2013cya,Liu:2014spa,Liu:2013vfa,Ketzer:2015tqa,Szczepaniak:2015eza,Guo:2015umn,Liu:2015taa,Liu:2015fea,Achasov:2015uua,Guo:2016bkl,Aceti:2012dj,Aceti:2016yeb,Bayar:2016ftu,Roca:2017bvy,Liu:2016onn,Liu:2016xly,Liu:2016dli,Liu:2015cah,Xie:2017mbe},
and a recent review of the TS mechanism can be found in
Ref.~\cite{Guo:2017jvc}.

As follows, in Sec. II we first introduce the TS mechanism in
\lamPKpi with a detailed analysis of the kinematic features in this
process, and an effective formulation for the transition amplitudes
is also constructed. Then, in Sec. III the numerical results are
presented. Finally, a brief summary is given in Sec. IV.

\section{The Model}

\begin{figure}[htb]
    \centering
    \includegraphics[width=0.95\linewidth]{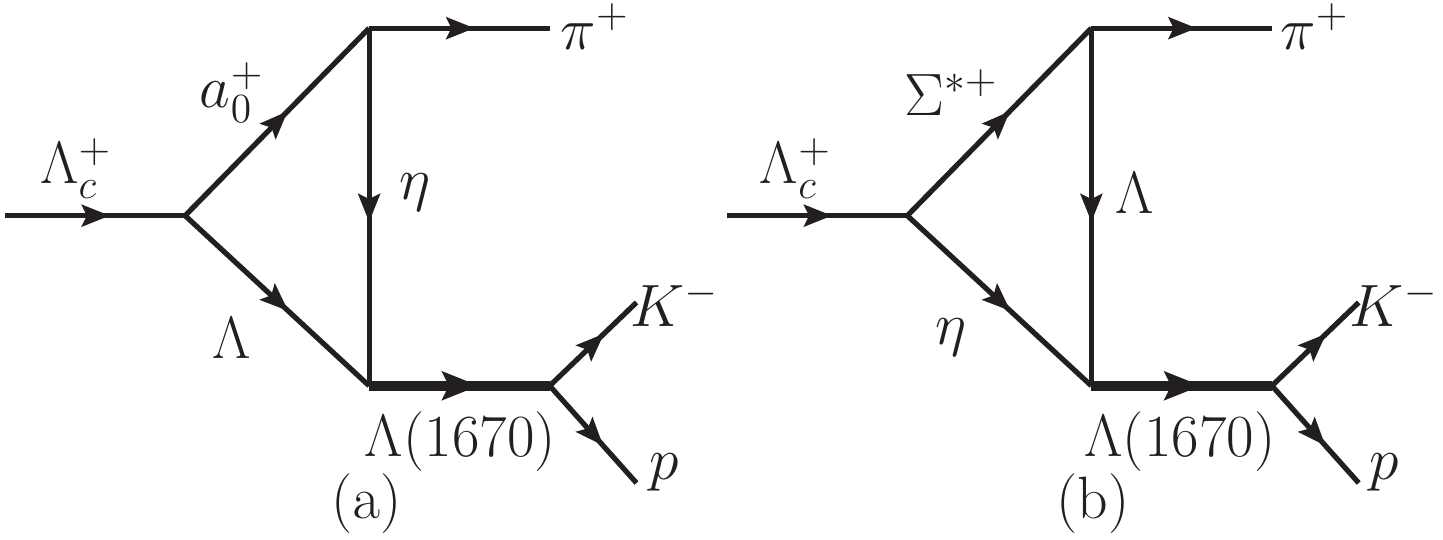}
    \caption{Rescattering diagrams which contribute to $\Lambda_c^+\to p K^- \pi^+$. Kinematic conventions for the intermediate states are (a) $\Lambda(q_1, m_1)$, $a_0(q_2,m_2)$, $\eta (q_3, m_3)$ and (b) $\eta(q_1, m_1)$, $\Sigma^*(q_2,m_2)$, $\Lambda (q_3, m_3)$.}
    \label{feynman-diagram}
\end{figure}

Considering the Cabibbo-favored weak decays $\Lambda_c^+\to \Lambda a_0(980)^+$ and $\Lambda_c^+\to \eta \Sigma^{*+}$, where $\Sigma^{*+}$ represents an excited hyperon with $I=1$, the TS processes for \lamPKpi are illustrated in Fig.~\ref{feynman-diagram}.

\begin{figure}[htb]
    \centering
    \includegraphics[width=0.95\linewidth]{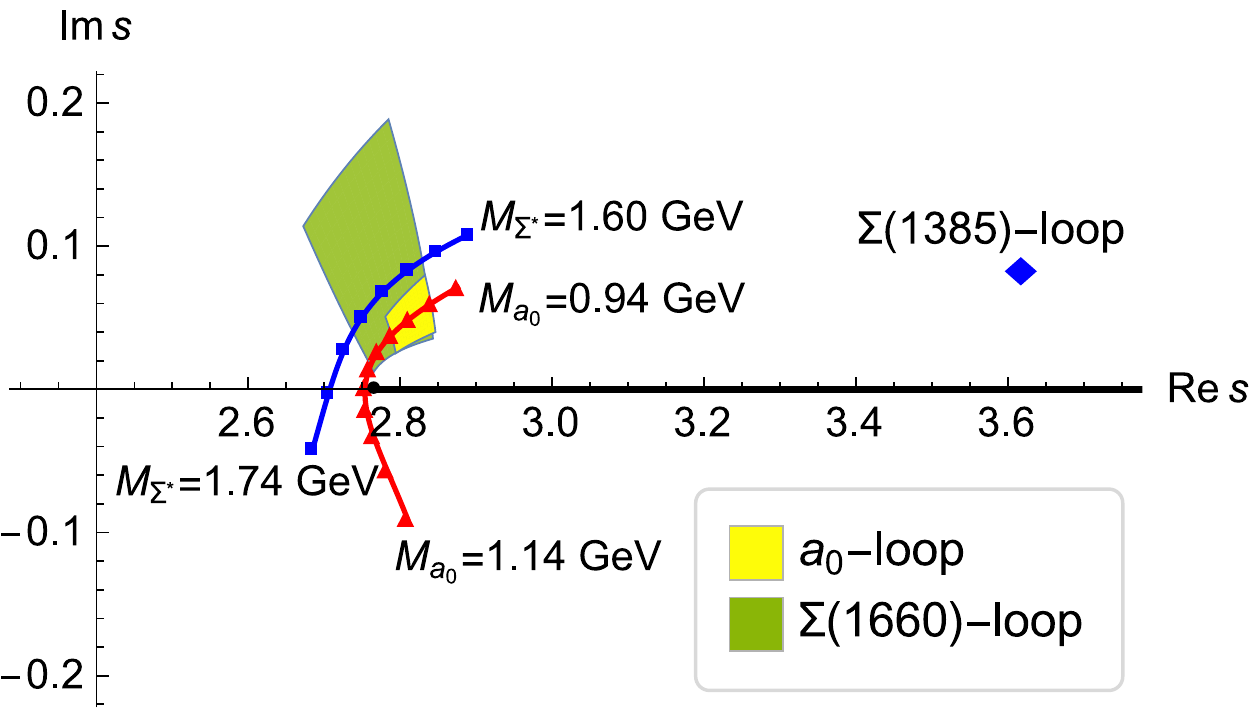}
    \caption{The TS location of $\mathcal{T}(s,m_2^2)$ in the complex $s$-plane. The thick line on the real axis represents the unitarity cut starting from $s_{\text{th}}$. The trajectory marked with triangle (box) is obtained by varying $M_{a_0}$ ($M_{\Sigma^*}$) and fixing $\Gamma_{a_0}=75\ \mbox{MeV}$ ($\Gamma_{\Sigma^*}=100\ \mbox{MeV}$).}
    \label{TS-position}
\end{figure}

We define the $K^-p$ invariant mass square
$s\equiv(p_{K^-}+p_p)^2\equiv M_{K^-p}^2$. Apparently the decay
amplitude $\mathcal{T}(s)$ corresponding to
Fig.~\ref{feynman-diagram}(a)/(b) has a normal threshold singularity
at $s_{\text{th}}\equiv (m_\Lambda +m_\eta)^2$, which is the start
point of the right-hand unitarity cut in the complex $s$-plane. This
cut results in a two-sheet structure for $\mathcal{T}(s)$, and the
physical region is just above the real axis $s\geq s_{\text{th}}$ on
the first Riemann sheet (RS), shown as the thick line in
Fig.~\ref{TS-position}. This unitarity cut leads to a cusp in the
$K^- p$ spectrum. In some special kinematical configurations, all of
the three intermediate states in Fig.~\ref{feynman-diagram}(a)/(b)
can be on-shell simultaneously, and the momenta of $\Lambda$ and
$\eta$ are parallel in the rest frame of $\Lambda_c^+$. In such a
case, the amplitude has a leading Landau singularity, which is
usually called the triangle singularity. The TS is found to be
located on the second RS
\cite{Bronzan:1963mby,Aitchison:1964zz,Schmid:1967ojm}. According to
the theorem of Coleman and Norton \cite{Coleman:1965xm}, the TS can
be present on the physical boundary (lower edge of the the second
RS) if, and only if, the triangle diagram can be interpreted as a
classical rescattering process in space-time. If the TS of a
rescattering amplitude is close to or just lies on the physical
boundary, it may result in a peak, or in another word, simulate a
resonance-like structure in the corresponding spectrum.

For the triangle diagram shown in Fig.~\ref{feynman-diagram}(a)/(b), the location of the TS in $s$ is given by \cite{Landau:1959fi,Coleman:1965xm,Bronzan:1963mby,Aitchison:1964zz}
\begin{eqnarray}\label{eq:TS}
s^{-}&=&(m_1+m_3)^2+\frac{1}{2m_2^2} {\LARGE[}(m_2^2+m_3^2-m_{\pi^+}^2) \nonumber \\  &\times&(M_{\Lambda_c^+}^2-m_1^2-m_2^2)-4m_2^2 m_1 m_3 \nonumber \\ &-& \lambda^{1/2}(M_{\Lambda_c^+}^2,  m_1^2,  m_2^2)\lambda^{1/2}(m_2^2,m_3^2,m_{\pi^+}^2){\LARGE ]},
\end{eqnarray}
with $\lambda(x,y,z)= (x-y-z)^2- 4yz$.

The TS is a logarithmic singularity. To avoid the infinity of the
loop integral in the physical region, one can replace the Feynman's
$i\epsilon$ for the $m_2$ propagator by $im_2 \Gamma_2$ with
$\Gamma_2$ the total decay width, or equivalently replace the real
mass $m_2$ by the complex mass
$m_2-i\Gamma_2/2$~\cite{Aitchison:1964ak}, which will remove the TS
from the physical boundary by a small distance if the width
$\Gamma_2$ is not very large, and the physical amplitude can still
feel the influence of this singularity. The physical meaning of this
prescription for avoiding the infinity is obvious: as long as the
kinematic conditions for the TS being present on the physical
boundary are fulfilled, it implies that the intermediate
state$^{[1]}$~\footnotetext[1]{Without causing ambiguities, we use
the mass symbols to represent the corresponding particles
somewhere.}$m_2$ is unstable, and it is necessary to take the
finite-width effects into account.  The above complex-mass scheme
provides a straightforward method to consistently implement the
unstable particles in calculating the amplitude. We refer to
Refs.~\cite{Denner:2014zga,Denner:2006ic,Du:2019idk} for more discussions about
the complex-mass scheme and
Refs.~\cite{Wu:2011yx,Wang:2013cya,Liu:2014spa,Liu:2013vfa,Guo:2014iya,Ketzer:2015tqa,Szczepaniak:2015eza,Guo:2015umn,Liu:2015taa,Liu:2015fea,Achasov:2015uua,Guo:2016bkl,Aceti:2012dj,Aceti:2016yeb,Bayar:2016ftu,Roca:2017bvy,Liu:2016onn,Liu:2016xly,Liu:2016dli,Liu:2015cah}
about the TS phenomena in various reactions.

In terms of Eq.~(\ref{eq:TS}),
with all the other masses fixed, the TS for the diagrams shown in Fig.~\ref{feynman-diagram} is on the physical boundary when $m_2^2$ falls in the range:
\begin{eqnarray}
\frac{m_1 m_{\pi^+}^2+ m_3 M_{\Lambda_c^+}^2}{m_1+m_3} -m_1 m_3\leq m_2^2\leq(M_{\Lambda_c^+}-m_1)^2,
\end{eqnarray}
corresponding to
\begin{eqnarray}
(m_1+m_3)^2 \leq s^- \leq(m_1+m_3)^2 +\frac{m_1[(m_2-m_3)^2-m_{\pi^+}^2] }{m_2}. \nonumber
\end{eqnarray}

Inputting the physical masses in Ref.~\cite{Tanabashi:2018oca}, the ranges for $m_2$ are $1.06\leq m_2\leq 1.17\ \text{GeV}$ and $1.70\leq m_2\leq 1.74\ \text{GeV}$, corresponding to Fig.~\ref{feynman-diagram}(a) and (b), respectively. One may notice that the mass of $a_0(980)$ is close to but does not falls in the above range. For Fig.~\ref{feynman-diagram}(b), there are several $\Sigma^{*}$ candidates, of which the masses are close to the above range, such as $\Sigma(1660)(J^P=1/2^+)$ and $\Sigma(1670)(J^P=3/2^-)$. But for the $\Sigma^*$ particle with higher spin, the $\Lambda_c^+\to \eta\Sigma^*$ decays proceed in higher partial waves. Considering that the $\eta\Sigma^*$ thresholds are close to $M_{\Lambda_c^+}$, those higher partial wave decays should be suppressed. For the diagram in Fig.~\ref{feynman-diagram}(b), we concentrate the discussion on the $\Sigma(1660)$-loop in this paper.
The $\Sigma(1660)$ is a three-star baryon cataloged in the PDG~\cite{Tanabashi:2018oca}, of which the mass and width are  approximately $1660$ MeV and $100$ MeV, respectively. Employing Eq.~(\ref{eq:TS}) and setting $m_2=M_{a_0,\Sigma^*}-i\Gamma_{a_0,\Sigma^*}/2$, the region where the TSs can appear in the $s$-plane is illustrated by the two colored blocks in Fig.~\ref{TS-position}, which are obtained by varying $M_{a_0}$ in the range of $960\sim 1000$ MeV, $\Gamma_{a_0}$ in $50\sim 100$ MeV, and $M_{\Sigma^*}$ in $1630\sim 1690$ MeV, $\Gamma_{\Sigma^*}$ in $40\sim 200$ MeV, respectively \cite{Tanabashi:2018oca}.

From Fig.~\ref{TS-position}, one can see that the TSs of both the $a_0(980)$-loop and $\Sigma(1660)$-loop are located on the upper half plane. Usually the singularities on the upper half plane of the second RS are supposed to be far from the physical region, the influence of which to the physical amplitude is negligible. However, we notice that the two TSs are still very close to the threshold in the sense of distance traveled in the complex plane, which means they can still affect the amplitude in the vicinity of $s_{\text{th}}$. The threshold $s_{\text{th}}$ is the closest point in the physical region to such TSs, and it is natural to expect that the cusp structure at $s_{\text{th}}$ could be enhanced by the nearby TS. This conclusion is numerically verified in the following.

Besides, the branching fraction of $\Lambda_c^+\to \Sigma(1385)^+ \eta$ is $(1.06\pm 0.32)\%$ \cite{Tanabashi:2018oca}, which is sizable.
But we find that the $\Sigma(1385)$-loop cannot result in the narrow peak in the $K^-p$ spectrum. For such a triangle diagram, the TS is far from the $\Lambda\eta$ threshold compared with the $a_0(980)$- or $\Sigma(1660)$-loop, which can be seen in Fig.~\ref{TS-position}. Only a less prominent cusp induced by the two-body unitarity cut can appear.

The general decay amplitude for $\Lambda_c^+\to \Lambda a_0(980)^+$ or $\Lambda_c^+\to \eta \Sigma(1660)^+$ can be written as
\begin{eqnarray}
\mathcal{M}(\Lambda_c^+\to \Lambda a_0^+ / \eta \Sigma^{*+})=g_A\bar{u}_f u_i + i g_B\bar{u}_f \gamma_5 u_i,
\end{eqnarray}
where $g_A$ and $g_B$ stand for the $S$- and $P$-wave couplings, respectively. There is no interference between the $S$- and $P$-wave amplitudes. We define the ratio $R\equiv |g_B|/|g_A|$, and find that the line-shapes of distribution curves are insensitive to the $R$ values. Therefore we set the $R$ to be a moderate value 1, and the numerical result is given in the unit of $|g_A|^2$.

{ The experimental data concerning $\Lambda_c^+\to \Lambda a_0^+$ or $\Lambda_c^+\to \eta \Sigma(1660)$  are not available yet.
Around half of the $\Lambda_c^+\to \Lambda\pi^+\eta$ decays are due to the two-body decays $\Lambda_c^+\to \Sigma(1385)^+ \eta$. Without taking into any interference, assuming that the $\Lambda_c^+\to \Lambda\pi^+\eta$ decays are saturated by the the resonant channels $\Sigma(1385)^+ \eta$ and $\Lambda a_0^+$ (or $\Sigma(1385)^+ \eta$ and $\Sigma(1660) \eta$), we can estimate the upper limit of $|g_A|^2$. Taking $Br(a_0^+ \to \eta \pi^+)\approx 1$ and $Br(\Sigma(1660) \to \Lambda \pi^+)\approx 0.128$ \cite{Kamano:2015hxa}, the upper limits are estimated to be $|g_A|^2_{\text{max}} \approx 0.32\ \text{GeV}^{-1}/\tau_{\Lambda_c}$ and $|g_A|^2_{\text{max}} \approx 3.14\ \text{GeV}^{-1}/\tau_{\Lambda_c}$  for the $\Lambda a_0^+$ and $\Sigma(1660) \eta$ channels, respectively.}

The $\eta\pi$ is the dominant decay channel of $a_0(980)$, and the pertinant amplitude reads
\begin{equation}
\mathcal{M}( a_0^+ \to \eta \pi^+)=g_{a_0\eta\pi}.
\end{equation}
For the strong decays $\mathcal{B}_i (\frac{1}{2}^+)\to \mathcal{B}_f (\frac{1}{2}^+) P$ and $\mathcal{B}_i (\frac{1}{2}^-)\to \mathcal{B}_f (\frac{1}{2}^+) P$, with $\mathcal{B}$ and $P$ indicating the baryon and light pesudoscalar meson respectively, the amplitudes take the forms
\begin{equation}
\mathcal{M}(\mathcal{B}_i\to \mathcal{B}_f P)=i g_{\mathcal{B}_i\mathcal{B}_f P} \bar{u}_f \gamma_5 u_i,
\end{equation}
and
\begin{equation}
\mathcal{M}(\mathcal{B}_i\to \mathcal{B}_f P)= g_{\mathcal{B}_i\mathcal{B}_f P} \bar{u}_f  u_i, \end{equation}
respectively. The coupling constant $g_{\mathcal{B}_i\mathcal{B}_f P}$ is determined by the pertinent partial decay width.
The $\eta\Lambda \to K^- p$ reaction is dominated by the $\Lambda(1670)$ ($J^P=1/2^-$) pole around the energy region we are interested in. Within the chiral unitary approach, the $\Lambda(1670)$ is supposed to be dynamically generated from the $S$-wave meson-baryon interactions in the strangeness $S=-1$ sector \cite{Oset:2001cn}.

The rescattering amplitude of \lamPKpi via the rescattering process is given by
\begin{eqnarray}\label{loop-amplitude}
&&\mathcal{T} = \frac{1}{s-M_{\Lambda(1670)}^2+iM_{\Lambda(1670)} \Gamma_{\Lambda(1670)}} \nonumber \\ && \times \int \frac{d^4q_1}{(2\pi)^4} \frac{\mathcal{A}  }{ (q_1^2-m_{1}^2) (q_2^2-m_{2}^2) (q_3^2-m_3^2) } ,
\end{eqnarray}
with $\mathcal{A}=\mathcal{M}(\Lambda_c^+\to \Lambda a_0^+)\mathcal{M}(a_0^+\to \eta \pi^+)\mathcal{M}(\eta\Lambda \to \Lambda(1670))\mathcal{M}(\Lambda(1670)\to K^- p)$ and $\mathcal{M}(\Lambda_c^+\to \eta \Sigma^{*+})\mathcal{M}(\Sigma^{*+}\to \Lambda \pi^+)\mathcal{M}(\eta\Lambda \to \Lambda(1670))\mathcal{M}(\Lambda(1670)\to K^- p)$ for $a_0(980)$-loop and $\Sigma(1660)$-loop, respectively, where the sum over polarizations of intermediate state is implicit.

For the $\Sigma(1660)$-loop, an additional regulator $\mathbb{F}(q_1^2)=(m_1^2-\Lambda_{\rm cut}^2)/(q_1^2-\Lambda_{\rm cut}^2)$ is introduced to kill the ultraviolet divergence that appears in the loop integral. This cutoff energy $\Lambda_{\rm cut}$ is a model-dependent parameter. However, it is found that when $\Lambda_{\rm cut}$ increases from 1 to 3 GeV, the variation of the distribution curve is very tiny. This can be qualitatively understood as the following: the dominant contribution to the loop integral in Eq.~(\ref{loop-amplitude}) comes from the momentum region where the intermediate particles are (nearly) on-shell, i.e. when $q_1^2=m_{1}^2$, $\mathbb{F}(q_1^2)$ gives 1; Furthermore, if we use extremely large values of $\Lambda_{\rm cut}$, the regulator $\mathbb{F}(q_1^2)$ will always be nearly one within a large integration interval \cite{Liu:2017vsf}.
Since the line-shape of the distribution curve is insensitive to the $\Lambda_{\rm cut}$ value, we only show the results by fixing $\Lambda_{\rm cut}$ at 2 GeV.

\section{Numerical results}

\begin{figure}[htb]
    \centering
    \includegraphics[width=1.0\linewidth]{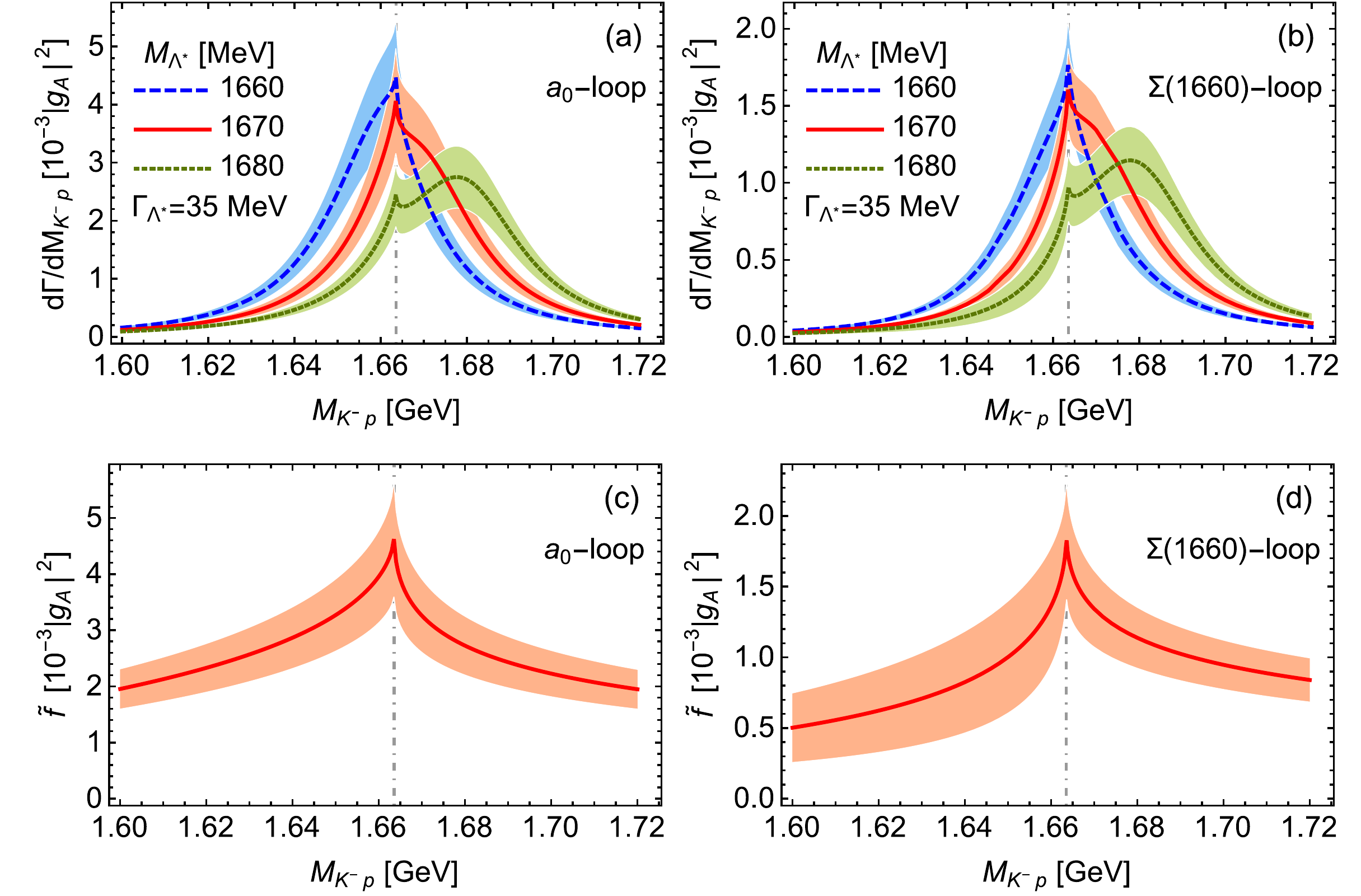}
    \caption{Invariant mass distribution of $K^-p$ via the rescattering processes in Fig.~\ref{feynman-diagram}. The bands are obtained by taking into account uncertainties of the mass and width of $a_0(980)$/$\Sigma(1660)$. The vertical dot-dashed line indicates the $\Lambda\eta$ threshold.}
    \label{pK-Spectrum-all}
\end{figure}

The $K^-p$ invariant mass distributions via Figs.~\ref{feynman-diagram}(a) and (b) are shown in Figs.~\ref{pK-Spectrum-all}(a) and (b), respectively. A narrow peak just staying at the $\Lambda\eta$ threshold--1663 MeV can be clearly seen in both of the two plots. This is the cusp structure enhanced and narrowed by the nearby TSs. The mass of $\Lambda(1670)$ is very close to the $\Lambda\eta$ threshold, and this cusp just grows up on the $\Lambda(1670)$ resonance bump in the $M_{K^-p}$ spectrum. Therefore the line-shape of the $\Lambda(1670)$ pole may disturb the identification of the narrow cusp.
The PDG gives that the mass of $\Lambda(1670)$ is in the range of 1660 to 1680 MeV, with the averaged value $M_{\Lambda^*}\approx 1670\ \mbox{MeV}$, and the width is in the range of 25 to 50 MeV, with the averaged value $\Gamma_{\Lambda^*}\approx 35\ \mbox{MeV}$ \cite{Tanabashi:2018oca}.  The dependence of $M_{K^-p}$ distribution curve on $M_{\Lambda^*}$ is illustrated in Figs.~\ref{pK-Spectrum-all}(a) and (b) by setting $M_{\Lambda^*}$ at 1660, 1670 and 1680 MeV, separately. Although the three curves in Fig.~\ref{pK-Spectrum-all}(a) or (b) behave differently, the peak position of the narrow cusp is not shifted.

In order to eliminate the influence of the $\Lambda(1670)$ pole in identifying the cusp, we can define a new distribution function
\begin{equation}
    \Tilde{f}(M_{K^-p})=\left|\frac{s-M_{\Lambda^*}^2+iM_{\Lambda^*} \Gamma_{\Lambda^*}}{M_{\Lambda^*}\Gamma_{\Lambda^*} } \right|^2 \times \frac{d\Gamma}{ d M_{K^-p}}.
\end{equation}
The corresponding distribution curves are displayed in Figs.~\ref{pK-Spectrum-all}(c) and (d), where we can see the narrow peaks at the $\Lambda\eta$ threshold still exist. This implies that even without introducing a genuine resonance, the cusp enhanced by the nearby TS can still simulate a narrow resonance-like structure.

The diagrams in Fig.~\ref{feynman-diagram} only account for the reactions which produce the signal--narrow cusp structure in \lamPKpi. The three-body decays \lamPKpi are dominated by the resonant subchannels $p\bar{K}^{*0}$, $\Delta^{++} K^-$ and $\Lambda(1520)\pi^+$, and a smooth nonresonant background \cite{Yang:2015ytm}. However, it can be seen that in Fig.~3 of Ref.~\cite{Yang:2015ytm}, the influence of these intermediate resonances can be well separated from the cusp structure by a proper cut in the Dalitz plot.

\begin{figure}[htb]
    \centering
    \includegraphics[width=0.9\linewidth]{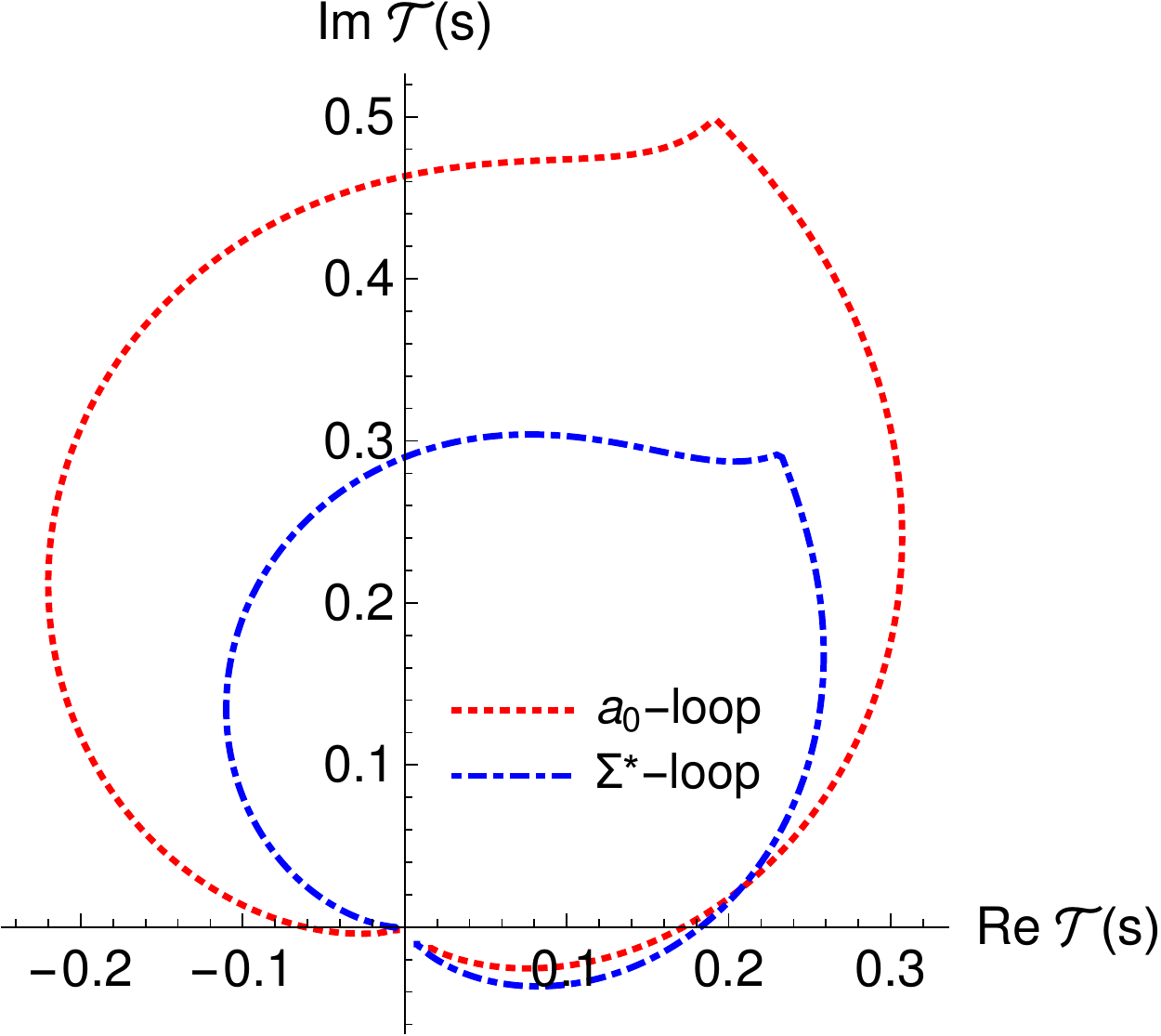}
    \caption{Argand plot of the rescattering amplitude, with $s$ increasing from $(M_{K^-}+m_p)^2$ to $(M_{\Lambda_c^+}-m_{\pi^+})^2$ counterclockwise. The mass and width of $\Lambda(1670)$ are taken to be PDG averaged values.}
    \label{argand}
\end{figure}

Since the line-shape of the distribution curve for a narrow cusp is similar to that for a genuine resonance pole, we need some criteria to distinguish these two underlying structures. One criterion is to check the difference between the Argand plots of corresponding amplitudes.
Taking the nominator $\mathcal{A}$ in Eq.~(\ref{loop-amplitude}) to be $-1$, the corresponding Argand plots of $\mathcal{T}(s)$ for the $a_0(980)$- and $\Sigma(1660)$-loop are shown in Fig.~\ref{argand}. We can see that neither of the plots is a perfect circle, however the plot should be for a genuine resonance. The peak on the imperfect circle reflects the rapid variation of the rescattering amplitude at the threshold and is correlated with the cusp in the invariant mass spectrum.

\section{Summary}

In summary, we investigate the \lamPKpi decays via the intermediate
$a_0(980)\Lambda$ and $\Sigma(1660) \eta$ rescattering processes,
and find that the narrow resonance-like structure observed by Belle
in the $K^-p$ invariant mass spectrum can be identified as the cusp
phenomenon caused by the $\Lambda\eta$ open threshold and closely
related to the TS mechanism$^{[2]}$\footnotetext[2]{It should be mentioned
that based on the analysis of the $K^- p \to \eta \Lambda$
reaction~\cite{Starostin:2001zz}, a $J^P = 3/2^-$ $\Lambda^*$ state
with mass around 1670 MeV and much narrow width was introduced in
Refs.~\cite{Liu:2011sw,Liu:2012ge,Liu:2012bk}. Later, a similar
state, but with $J^P = 3/2^+$, was further investigated in
Refs.~\cite{Kamano:2014zba,Kamano:2015hxa} with a dynamical
coupled-channels model. Possibly these theories can also explain the
observations of Belle by adjusting some parameters, but the above
quantum number assignments of this narrow structure are different
from that in our scenario. Further partial wave analysis of the
experimental data is desirable to confirm or rule out some of these
interpretations.}. Such a special phenomenon is due to the
analytical property of the scattering amplitudes with the TS located
to the vicinity of the physical boundary. This will enhance the
two-body cusp effect and make it more predominant than the usual cases.
In addition, we show that the TS enhanced cusp structure can mimic a resonance behavior in the Argand diagram. But with sufficiently high luminosity one may still be able to measure effects from the TS and cusp mechanism, and distinguish the TS enhanced cusp structure from a genuine resonance. Experiments at BESIII, Belle-II, and LHCb should have advantages of probing such a mechanism.
As a direct prediction of the proposed mechanism, since the $\Lambda\eta$ channel also strongly couples to $\Sigma\pi$ around 1670 MeV, we anticipate that a similar cusp structure can also be observed in $\Lambda_c\to \Sigma\pi\pi$. Future experiments or analyses at BESIII, Belle, Belle-II, and LHCb can provide a test of this scenario.

\begin{acknowledgments}

Helpful discussions with Bing-Song Zou and Cheng-Ping Shen are gratefully acknowledged. This work is supported, in part, by the National Natural Science Foundation of China (NSFC) under Grant Nos.~11425525, 11521505, 11675091, 11735003, 11835015
and 11475227, by the Deutsche Forschungsgemeinschaft (DFG) and NSFC through funds provided to the Sino--German Collaborative Research Center
``Symmetries and the Emergence of Structure in QCD'' (NSFC Grant
No.~11621131001, DFG Grant No.~TRR110), and by the National Key Basic Research Program of China under Contract No.~2015CB856700. It is also partly supported
by the Youth Innovation Promotion Association CAS (No. 2016367).

\end{acknowledgments}

\end{document}